\documentclass[11pt,a4paper]{article}
\pdfoutput=1
\usepackage{bbm}
\usepackage{amsmath}
\usepackage{stackrel}
\usepackage[pdftex]{hyperref}
\usepackage{float}
\usepackage{graphicx}

\newcommand{\be}{\begin{equation}}
\newcommand{\ee}{\end{equation}}
\newcommand{\ba}{\begin{eqnarray}}
\newcommand{\ea}{\end{eqnarray}}
\renewcommand{\(}{\left(}
\renewcommand{\)}{\right)}

\newcommand{\lk}{\left[}
\newcommand{\rk}{\right]}

\newcommand{\w}{\omega}
\newcommand{\hw}{\hat{\omega}}
\newcommand{\hq}{\hat{q}}
\newcommand{\nn}{\nonumber}
\newcommand{\g}{\gamma}

\begin{document}

\begin{flushright}
TUW-13-08\\
\end{flushright}

\vskip 2cm \centerline{\Large {\bf Holographic thermalization in $\mathcal{N}=4$ Super  Yang-Mills}}
\vskip 0.3cm
\centerline{\Large{\bf theory at finite coupling}}
\vskip 2cm
\renewcommand{\thefootnote}{\fnsymbol{footnote}}
\centerline
{{\bf Stefan A. Stricker
\footnote{stricker@hep.itp.tuwien.ac.at}
}}
\vskip .5cm
\centerline{\it Institut f\"{u}r Theoretische Physik, Technische Universit\"{a}t Wien}
\centerline{\it Wiedner Hauptstr.~8-10, A-1040 Vienna, Austria}

\setcounter{footnote}{0}
\renewcommand{\thefootnote}{\arabic{footnote}}

\begin{abstract}
We investigate the behavior of   energy momentum tensor correlators  in holographic $\mathcal{N}=4$ super Yang-Mills plasma, taking finite coupling corrections into account. In the thermal limit we determine the flow of  quasinormal modes as  a function of the 't Hooft coupling. 
 Then we use a specific model of holographic thermalization to study the deviation of the spectral densities from their thermal limit in an out-of-equilibrium situation.
 The main focus lies on the thermalization pattern with which the plasma constituents approach their thermal distribution as the coupling constant decreases from the infinite coupling limit.
All obtained results point towards the weakening of the usual top-down thermalization pattern.
 \end{abstract}

\newpage
\tableofcontents

\section{Introduction}

Understanding the complicated field dynamics in a heavy ion collision presents a difficult challenge to QCD theorists.
Experiments at RHIC and the LHC point  towards the conclusion that the quark gluon plasma (QGP)  created in heavy ion collisions  behaves as a strongly coupled, nearly perfect,  liquid \cite{Tannenbaum:2012ma, Muller:2012zq} rather than a weakly interacting gas of quarks and gluons.
The strongly coupled nature of the created matter has made the gauge/gravity duality one of the standard tools in describing QGP physics \cite{DeWolfe:2013cua, CasalderreySolana:2011us}, supplementing traditional approaches such as perturbation theory or lattice gauge theory.

In its original form the gauge gravity duality relates supergravity on five-dimensional  asymptotically anti deSitter space time (AdS)  to  strongly  coupled $\mathcal{N}=4$ super Yang Mills (SYM) theory living on the boundary of the AdS space.
Although SYM is  very different from QCD in its vacuum state, it shares many features with QCD in the deconfined phase, such as a finite screening length, Debye screening and broken supersymmetry.

One particularly useful development is the application of the duality to out-of-equilibrium systems 
by mapping the thermalization process to black hole formation in asymptotically AdS space.
This has  led to the insight that fluid dynamics becomes a good approximation rather quickly, but this does not mean that the system is isotropic or thermal  \cite{Chesler:2008hg, Chesler:2010bi, Heller:2013fn, Wu:2011yd, Heller:2012km}.

A particularly important challenge in an out-of-equilibrium system is to identify the thermalization pattern with which the  plasma constituents of different energies approach their thermal distribution.
On the weakly coupled side classical calculations have shown that the thermalization process is of the bottom-up type, i.e. low energetic modes reach thermal equilibrium first, with inelastic scattering processes being the driving mechanism behind it \cite{Baier:2000sb}.
In the early stages many soft gluons are emitted which  form a thermal bath very quickly and then draw  energy from the hard modes.
 Recently this picture got supported by numerical simulations \cite{Berges:2013eia}. In \cite{Kurkela:2011ti} an alternative proposal was made: the thermalization process is driven by instabilities which isotropize the momentum distributions more rapidly than scattering processes \footnote{See also \cite{Attems:2012js}}. 
 On the contrary, holographic calculations in the infinite coupling limit always point towards top-down thermalization, where the high energetic modes reach equilibrium first, indicating a probable transition between the two behaviours at intermediate coupling \cite{Balasubramanian:2010ce, CaronHuot:2011dr, Galante:2012pv, Erdmenger:2012xu}.

Evaluating non-local observables such as two point functions in a time dependent thermalizing system is an extremely challenging but important  task, since they allow to see how different energy/length scales approach thermal equilibrium. 
One strategy was worked out in  \cite{CaronHuot:2011dr} where it was shown how  fluctuations  created near the horizon and dissipation come to a balance to satisfy a generalized fluctuation dissipation theorem.
 For a different approach to generalize the fluctuation dissipation theorem see \cite{Mukhopadhyay:2012hv}. 
 Non equilibrium  generalizations of spectral functions and occupation numbers were introduced in  \cite{Banerjee:2012uq, Balasubramanian:2012tu}.
 Complementary quantities of interest are entanglement entropy   and Wilson loops \cite{Ryu:2006bv, Hubeny:2007xt}.

A particularly useful model, due to its simplicity is the collapsing shell model, where the thermalization process is mapped to the collapse of a spherical shell of matter and the subsequent formation of  a black hole \cite{Danielsson:1999fa, Danielsson:1999zt, Ebrahim:2010ra, Wu:2012rib, Wu:2013qi, Garfinkle:2011tc, Baron:2012fv, Galante:2012pv, Aparicio:2011zy, Keranen:2011xs, Lin:2013sga, Zeng:2013mca, Caceres:2012em, Caceres:2012px, Caceres:2013dma}. 
In the limit where the shell's motion is slow compared to the other scales of interest this model was used  to study the approach of the spectral density to  equilibrium  for the components of the energy momentum tensor in \cite{Lin:2008rw} and for dileptons and photons in \cite{Baier:2012tc, Baier:2012ax}\footnote{For the effect of anisotropies on the photon production and shear viscosity see \cite{Rebhan:2011vd, Arciniega:2013dqa, Patino:2012py, Wu:2013qja, Giataganas:2013lga}}.
 All these studies show the usual top-down thermalization pattern.
In addition in \cite{Steineder:2013ana} the virtuality dependence of the photons was taken into account, showing that on-shell photons are last to thermalize, consistent with the conclusions from other models of holographic thermalization \cite{Chesler:2012zk, Arnold:2011qi}.

Due to its simplicity the collapsing shell model, in the quasi-static limit, even allows one to include the first order string corrections to the supergravity action and leave the infinite coupling limit. 
In \cite{Steineder:2012si} the leading order string corrections to the photon spectral density \cite{Hassanain:2011ce, Hassanain:2012uj} were generalized   to an out-of-equilibrium situation, 
showing indications that the usual top-down thermalization pattern shifts towards bottom-up when finite coupling corrections are included.
This observation was strengthened by a quasinormal mode (QNM) analysis at finite coupling in \cite{Steineder:2013ana}. As the coupling constant is decreased the tower of poles bends towards the real axis, also showing a weakening of the top-down pattern, but this time independent of the thermalization model being used.
It still needs to be investigated if the observed change is intrinsic to photons or/and  a consequence of the collapsing shell model or of more general validity.

In \cite{Baron:2013cya} the AdS-Vaidya solution  was used to investigate the 
thermalization time scale for non-local observables in SU(N) $\mathcal{N}=4$ SYM theory at finite coupling using geometric probes in the bulk.
 Interestingly, there the UV modes thermalize faster and the IR modes slower if the coupling constant is decreased from the infinite coupling limit.
 The authors speculate that the difference between their analysis and the one for photons \cite{Steineder:2012si, Steineder:2013ana} originates from the fact that in order to study current correlators, it is necessary to include the Ramond-Ramond five form field strength   in the $\mathcal{O}(\alpha'^3)$ corrections, which produce very large corrections to observables associated with electric charge transport. We will say more about this  in the conclusions.

The goal of this paper is to shed light on the above issues  by studying  energy momentum tensor correlators of  a  $\mathcal{N}=4$ SYM plasma and their approach to thermal equilibrium at finite coupling in the collapsing shell model.
In the infinite coupling limit the correlators were first studied in \cite{Kovtun:2006pf, Teaney:2006nc, Kovtun:2005ev} and to next-to-leading order in a strong coupling expansion in \cite{Buchel:2004di, Benincasa:2005qc}. The leading order corrections in  inverse powers  of numbers of colours, $N_c$,  was computed in \cite{Myers:2008yi, Buchel:2008ae}.
Finite coupling effects on jet quenching were worked out in \cite{Arnold:2012uc, Arnold:2012qg}.
The out-of-equilibrium dynamics using the collapsing shell model  at infinite coupling was considered in \cite{Lin:2008rw}.
In the paper at hand   we  fill the missing gap by analyzing the flow of the quasinormal mode spectrum as a  function of the coupling constant as well as the  approach of the spectral density to its thermal value at finite coupling  in the collapsing shell model. 

The paper is organized as follows. In section \ref{setup} we  will review the collapsing shell model and introduce the finite coupling corrections. After that we outline the main parts of the calculation in section \ref{sec:spectral} and present the results for the quasinormal modes and the spectral densities in section \ref{results}. In section \ref{conclusion} we draw our conclusions.

\section{Setup}\label{setup}
\subsection{The collapsing shell model}

Our aim is to use the collapsing shell model introduced in  \cite{Danielsson:1999fa, Danielsson:1999zt} to gain insights into the  thermalization process of a strongly coupled $\mathcal{N}=4$ SYM plasma via the gravitational collapse of a spherically symmetric  shell of matter in anti deSitter space.
On the field theory this corresponds to the preparation of an exciting state through the injection of energy and the subsequent evolution towards thermal equilibrium.

Following Birkhoff's theorem,  outside the shell the background is given by a black hole solution, whereas inside the shell the metric is given by its zero temperature counterpart.
The five-dimensional AdS metric is given by
\be\label{AdS5}
ds^2=\frac{r_h^2}{L^2 u}\lk f(u)dt^2+dx^2+dy^2+dz^2\rk+\frac{L^2}{4 u^2f(u)}du^2\;,
\ee
where
\be\label{f}
f(u) \,=\, \left\{ \begin{array}{lr}
f_+(u)=1- u^2 \, ,& \mathrm{for}\;u >u_s\\
f_-(u)=1\, ,& \mathrm{for}\; u < u_s
\end{array}\right. \;,
\ee
and $u\equiv r_h^2/r^2$ is a dimensionless coordinate where the boundary is located at $u=0$ and the horizon at $u=1$. From now on the index '--' denotes the inside and '+' the outside 
 space time of the shell and  we set the curvature radius of  $AdS$  to $L=1$. 
 
The shell can be described by the action for a membrane \cite{Lin:2008rw}
\be
S_m=-p \int d^4\sigma \sqrt{-\det g_{ij}}\;,
\ee
where $g_{ij}$ is the induced metric on the brane and $p$ is the only parameter that characterizes the membrane.
Due to the discontinuity of the time coordinate in the above metric, fields living in the above background have to be matched across the shell using the Israel matching conditions given by 
\be
[K_{ij}]=\frac{\kappa_5^2 p}{3} g_{ij}\;,
\ee
where $[K_{ij}]=K^+_{ij}-K^-_{ij}$ is the extrinsic curvature and $\kappa_5^2=8\pi G_5$ is Newtons constant in the Einstein frame \footnote{ In our numerical calculations  we always  set   $\kappa_5^2p=1$. 
}.
 From the above equation   the trajectory of the shell is determined. 

Physical initial conditions for the shell,  that could be a good approximation  for heavy ion collisions,  are determined through the relation  of the holographic coordinate with the temperature $r_h=T \pi$ and the saturation scale $r_s=Q_s\pi$ together with a vanishing initial velocity \cite{Lin:2013sga}.
For  LHC these values are $T\sim 400$ MeV and $Q_s\sim1.23$ GeV.

We, however, are not going to treat the  dynamical process but
 work in the quasi-static approximation, where the motion of the shell is slow compared to the other scales of interest and only take snapshots of the shell at certain positions. See  appendix \ref{A} for the exact relation when the quasi-static approximation holds. 
 This condition, however, breaks down at the latest stages of the collapse as can be seen by comparing  
 the Penrose diagram for the black hole space time with the collapsing shell diagram \cite{Lin:2013sga}.

When the quasi-static approximation is applicable, the matching conditions simplify considerably and   explicit calculations in frequency space are possible.
In this case the discontinuity of the time coordinate implies that the frequencies measured inside and outside the shell are related through \cite{Lin:2008rw}
\be\label{omega}
\omega_-=\frac{\omega_+}{\sqrt{f_s}}=\frac{\omega}{\sqrt{f_s}}\;, \qquad f_s=f_+(u_s).
\ee
The subscript $s$ denotes the position of the shell at $u=u_s$.
The matching conditions at the shell  for metric perturbations of the form
\be
g_{\mu\nu}\rightarrow g_{\mu\nu}+h_{\mu\nu}\,,
\ee 
 have been worked out for all the relevant components in \cite{Lin:2008rw}.
For example for the $xy$ component they are
 \ba\label{mcxy}
  h^-_{xy}\big\vert_{u_s}&=& \sqrt{f_s}h^+_{xy}\big\vert_{u_s},\\
  \partial_u h^-_{xy}+\frac{2\kappa_5^2p}{3u}h^-_{xy}\big\vert_{u_s}&=& f_s \partial_u h^+_{xy}\big\vert_{u_s}.
 \ea

\subsection{Finite coupling corrections }

In order to leave the strict $\lambda=\infty$ limit the  leading order string corrections to type IIB supergravity have to be included and this is accounted for by the action \cite{Gross:1986iv, Gubser:1998nz}
\be\label{action}
S_{IIB}=\frac{1}{2\kappa_{10}}\int d^{10}x\sqrt{-g}\lk R_{10}-\frac{1}{2}(\partial \phi)^2-\frac{1}{4.5!}(F_5)^2+\cdot\cdot\cdot+\gamma \,e^{-\frac{3}{2}\phi}C^4\rk
\ee
where $\gamma\equiv\frac{1}{8}\zeta(3)\lambda^{-\frac{3}{2}}$. $F_5$ is the five form field strength and $\phi$ is the dilaton.
The $C^4$ term is proportional to the fourth power of the Weyl tensor
\be
C^4=C_{hmnk}C_{pmnq}C_h^{\;rsp}C^q_{\;rsk}+\frac{1}{2}C^{hkmn}C_{pqmn}C_h^{\;rsp}C^q_{\;rsk}.
\ee 
The $\gamma$-corrected AdS black hole metric derived from the above action can be written as \cite{Gubser:1998nz,Pawelczyk:1998pb,Paulos:2008tn}
\ba\label{AdSg}
ds^2_{10}&=& g_{5mn}dx^mdx^n+c_4 d\Omega_5^2\nn\\
&=&-c_1dt^2+c_2 d\mathbf{x}^2 +c_3 du^2+c_4 d\Omega_5^2,
\ea
where the coefficients $c_i=c_i(u)$ only  depend on the dimensionless holographic  coordinate $u=r_h^2/r^2$ and $d\Omega_5^2$ is the metric of a  five-dimensional  unit sphere.
The solution can be written explicitly as
\ba
c_1 &=&\frac{r_h^2}{u} f(u) e^{a(u)-\frac{10}{3}\nu(u)}\;,\nn\\
c_2&= &\frac{r_h^2}{u}  e^{-\frac{10}{3}\nu(u)}\;,\\
c_3&= &\frac{1}{4 u f(u)}  e^{b(u)-\frac{10}{3}\nu(u)}\nn\\
c_4&=&e^{2\nu(u)}\nn
\ea
with $f(u)$ given in (\ref{f}) and
\ba
a(u) &=& -15 \gamma (5u^2+5u^4-3u^6)\;,\nn\\
b(u)&= &15\gamma (5u^2+5u^4-19u^6)\;,\\
 \nu(u)&=&\gamma \frac{15}{32}u^4(1+u^2).\nn
\ea
The $\gamma$-corrected relation between $r_h$ and the field theory temperature reads $r_h=\pi T/(1+15 \gamma)$.
Note that the vacuum solution of AdS does not receive $\gamma$-corrections \cite{Banks:1998nr}.
In the next section we will calculate the spectral densities for the different symmetry channels  of the
energy momentum tensor obtained from the above metric.

\section{Spectral density }\label{sec:spectral}
In order to see how the plasma constituents  approach thermal equilibrium we  study
the spectral densities of various energy momentum tensor components by considering  linearized perturbations of the five dimensional metric 
\be
g_{\mu\nu}\rightarrow g_{\mu\nu}+h_{\mu\nu},
\ee
where the linear perturbations $h_{\mu\nu}$ correspond to the energy momentum tensor of the field theory.
Following  \cite{Kovtun:2005ev} the metric perturbations can be combined into  three gauge invariant fields $Z_s$ representing the three symmetry channels, namely  spin 0 (sound channel), spin 1 (shear channel) and  spin 2 (scalar channel). 

Going to momentum space we look at fluctuations of the form
 \be
 h_{\mu\nu}(t,\mathbf{x},u)=\int \frac{d^4 k}{(2\pi)^4}e^{-i \omega t+i \mathbf{q}\mathbf{x}}h_{\mu\nu}(u)\;.
 \ee
 In the following it will be convenient to introduce the dimensionless quantities 
 \be
 \hw=\frac{\w}{2\pi T}\,,\qquad \hq=\frac{q}{2\pi T}.
 \ee

Deriving the equations of motion for the different symmetry channels is a lengthy and tedious exercise which has been performed   in the literature before \cite{Benincasa:2005qc, Buchel:2004di, Buchel:2008bz}.
Therefore we  shall only describe the main points of the derivation here and guide the interested reader to the relevant references for further details.
We will extend the  solutions obtained in the hydrodynamic limit \cite{Benincasa:2005qc, Buchel:2004di, Buchel:2008bz} to arbitrary momenta and energies.

\subsection{Scalar channel}

The EoM for the  scalar channel  is obtained by considering the metric fluctuations $h_{xy}$.
It is convenient to introduce a field
\be
Z_1=g^{xx}h_{xy}=\frac{u}{r_h^2}h_{xy}.
\ee
By expanding this field to linear order in $\gamma$
\be
Z_1=Z_{1,0}+\g Z_{1,1}+\mathcal{O}(\g^2)
\ee
the EoM  for the scalar channel takes the compact form  \cite{Buchel:2008bz}
\ba\label{eomscalar}
Z_1 ''&-&\frac{u^2+1}{uf}Z_1'+\frac{\w^2-\hq^2f}{u f^2}Z_1=-\frac{1}{4}\g \Bigg[ \(3171u^4+3840\hq^2u^3+2306u^2-600\)u\,Z_{1,0}' \nn\\
&+&\bigg(\frac{u}{f^2}\Big(600\hw^2-300\hq^2+50u+(3456\hq^2-2856\hw^2)u^2+768u^3\hq^4+(2136\hw^2-6560\hq^2)u^4\nn\\
&-&(768\hq^4+275)u^5+3404\hq^2u^6+225u^7 \Big)-30\frac{\hw^2-\hq^2f}{uf^2}\bigg) Z_{1,0}
\Bigg],
\ea
where the right hand side only depends on the zeroth order solution.
Note that we have an additional term of $\mathcal{O}(\gamma)$ (the last term) in the above equation compared to \cite{Buchel:2008bz}. This comes from the fact that we defined the dimensionless quantities $\hw$ and $\hq$ with respect to the $\gamma$-corrected temperature $T$ and not $T_0=r_h/\pi$ as in \cite{Buchel:2008bz}.

\subsection{Shear channel}
Following \cite{Kovtun:2005ev, Benincasa:2005qc},   the shear channel is defined by the metric fluctuations 
\be
\{h_{tx},h_{zx},h_{xu}\}.
\ee
Using the gauge condition $h_{xu}=0$ and introducing $H_{tx}=u h_{tx}/(\pi T)^2$ and  $H_{xz}=u h_{xz}/(\pi T)^2$ one can define the shear channel gauge invariant combination 
\be
Z_2=q H_{tx}+\w H_{xz},
\ee
for which one obtains a decoupled second order differential equation for $Z_2$ to $\mathcal{O}(\gamma)$ upon introducing
\be
Z_2=Z_{2,0}+\g Z_{2,1}+\mathcal{O}(\g^2).
\ee
 The equation of motion for the shear channel to $\mathcal{O}(\gamma)$ is given by
\be\label{eomshear}
Z_{2}''+\frac{1+u^2}{uf}Z_{2}'+\frac{\hw-\hq f}{uf^2}Z_{2}+\gamma J_2(Z_{2,0})\frac{u^2}{f}+\g\,\frac{30(\hw-\hq f)}{u f^2}Z_{2,0}=0,
\ee
The source term $\gamma J_2$ is of $\mathcal{O}(\gamma)$ and  depends only on the zeroth order solution $Z_{2,0}$ and derivatives thereof. The explicit form of this lengthy expression can be found in \cite{Benincasa:2005qc}. 
\subsection{Sound channel}
In order to investigate the sound channel we look at  metric perturbations of the form
\be
\{ h_{tt},\, h_{tz},\,h,\; h_{zz},\, h_{uu},\, h_{tu},\, h_{zu} \},
\ee
where $h=\sum _{\alpha=x,y} h_{\alpha\alpha}$ is a singlet.
After the equations of motion have been derived for the above perturbations the gauge conditions
\be
t_{tu}=h_{zu}=h_{uu}=0
\ee
are imposed.
In the sound channel there is a subtlety in constructing the gauge invariants due to the non constant warp factor in front of the unit five sphere. See \cite{Benincasa:2005iv, Benincasa:2005qc} for a detailed analysis of this issue.
It turns out that after introducing $\hat{h}_{tt}=\hat{c}_1H_{tt},\;\hat{h}_{tz}=\hat{c}_2H_{tz},\,\hat{h}=\hat{c}_2H$ and $\hat{h}_{zz}=\hat{c}_2H_{zz}$ 
the gauge invariant for the sound channel is given by
\be
Z_3=4\frac{q}{\w}H_{tz}+2 H_{zz}-H \(1-\frac{q^2}{\w^2}\frac{\hat{c}_1'\hat{c}_1}{\hat{c}_2'\hat{c}_2}\)+2 \frac{q^2}{\w^2}\frac{\hat{c}_1^2}{\hat{c}_2^2}H_{tt}\;,
\ee
where $\hat{c}_i=c_4^{5/3}c_i$.
Introducing 
\be
Z_3=Z_{3,0}+\g Z_{3,1}+\mathcal{O}(\g^2),
\ee
the equation of motion for this gauge invariant takes the form \cite{Benincasa:2005qc, Buchel:2008bz}
 \ba\label{eomsound}
Z_3''&-& \frac{\hq^2(2u^2-3-3u^4)+3(1+u^2)\hw^2}{uf(\hq^2(u^2-3)+3\hw^2)}Z_3' \nn\\
&+&\frac{\hq^4(3-4u^2+u^4)+3\hw^4+\hq^2(-6\hw^2+4u^2(u^3-u+\hw^2))}{uf^2(\hq^2(u^2-3)+3\hw^2)}Z_3\nn\\
&-& 
\gamma J_3(Z_{3,0})\frac{u^2}{f}+\g\frac{30(\hw-\hq f)}{u f^2}Z_{3,0}=0\;.
 \ea 
 Again, we are not showing the lengthy expression for the source term $J_3$ which only depends on the zeroth order solution and can be found in \cite{Benincasa:2005qc, Buchel:2008bz}. As before,  there is an additional term appearing in the equation of motion due to the different convention of the dimensionless quantities $\hw$ and $\hq$.
 
\subsection{Solving the EoMs}
We are now going to solve the equations of motion and determine the corresponding spectral densities. 
The equations  (\ref{eomscalar}), (\ref{eomshear}) and (\ref{eomsound})  have singular points at $u=\pm 1,0$. In the near horizon limit, $u\rightarrow 1$, the indicial exponents are given by $\mp i\hw/2$, where the minus sign corresponds to the infalling mode and the plus sign to the outgoing mode.
In thermal equilibrium, i.e. in the black hole background, one chooses the infalling boundary condition because classically nothing can escape from a black hole. However, in the presence of a shell the solution is a linear combination of  the ingoing and outgoing modes
\be
Z_{s,+}=c_-Z_{s,\mathrm{in}}+c_+Z_{s,\mathrm{out}},
\ee
where $s=1,2,3$ and the  coefficients $c_\pm$ are determined by the  matching conditions specified below.
In order to solve the EoMs (\ref{eomscalar}), (\ref{eomshear}) and (\ref{eomsound}) numerically we make the following ansatz for the ingoing and outgoing modes
\be
Z_{s,\substack{\mathrm{in}\\\mathrm{out}}}(u)=(1-u)^{\pm\frac{i\hw}{2}}\(Z^{(0)}_{s,\substack{\mathrm{in}\\\mathrm{out}}}(u)+\gamma Z^{(1)}_{s,\substack{\mathrm{in}\\\mathrm{out}}}(u)+\mathcal{O}(\gamma^2)\) \;,
\ee
where  $Z^{(0,1)}_{s,\mathrm{in},\mathrm{out}}$ is regular at the horizon $u=1$ and normalized to $Z^{(0,1)}_{s,\mathrm{in},\mathrm{out}}(u=1)=1$. 
We then integrate numerically from the horizon to the boundary.

This solution has to be matched to the zero temperature solution inside the shell.
At zero temperature there are no $\g$-corrections at leading order and therefore we can make use of the results obtained in \cite{Lin:2008rw}, where it was shown that the inside solution for all channels is given in terms of the Hankel function of the first kind
\be\label{hankel}
Z_{s,-}(u)= u H_1^{(2)}\( 2 \sqrt{u}\(\frac{\hw}{\sqrt{f^\g_s}}-\hq\)\).
\ee
The factor $f^\g_s$ enters by matching the inside frequency to the outside frequency via (\ref{omega})
\be\label{fmg}
\omega_-=\frac{\omega}{\sqrt{f^\g_s}},\qquad f^\g_s=f(u_s)e^{a(u_s)-\frac{10}{3}\nu (u)},
\ee
and is the only source of $\gamma$-corrections for the inside solution, which thus takes the form
\be
Z_{s,-}(u)=Z^{(0)}_{s,-}(u)+\gamma Z_{s,-}^{(1)}+\mathcal{O}(\g^2),
\ee
and can be obtained by a simple expansion of the Hankel function given in (\ref{hankel}).

The matching conditions for all channels have the compact form \cite{Lin:2008rw}
\begin{subequations}
\begin{align}
Z_{s,+},&=\frac{1}{\sqrt{f^\gamma_s}}Z_{s,-},\\
Z'_{s,+},&=\frac{1}{f^\g_s}Z'_{s,+}+\frac{1}{u}\(\frac{1}{\sqrt{f_s^\gamma}}+\frac{\kappa_5^2p}{3f_s^\gamma}\)Z_{s,+},
\end{align}
\end{subequations}
and lead to the ratio
\begin{eqnarray}\label{Cmp}
\frac{c_-(u_s)}{c_+(u_s)} = - \frac{Z_{s,in}\partial_u Z_{s,-}-\sqrt{f^\g_s} Z_{s,-} \partial_u Z_{s,in}}
{Z_{s,out}\partial_u Z_{s,-}-{\sqrt{f^\g_s}} Z_{s,-}\partial_u Z_{s,out}}\Bigg|_{u=u_s} =C_0+\gamma C_1+\mathcal{O}(\gamma^2)\;,
\end{eqnarray}
where all the $\gamma$-corrections have to be taken into account.
This ratio, in particular the parametric dependence of $C_0$ and $C_1$ on the frequency,   will play an important role for the thermalization pattern.
Having solved for the ingoing and outgoing modes the
retarded correlators for the gauge invariants can be calculated via 
 standard AdS/CFT techniques  \cite{Teaney:2006nc, Kovtun:2006pf}, which produce

\ba\label{spectral}
G_s(\hw,\hq,u_s,\g))&=&-\pi^2 N_c^2 T^4\(1-\frac{15}{2}\g\) \(\frac{Z''_{s,+}}{2Z_{s,+}}\) \Bigg|_{u=0}\nn\\
&=&-\pi^2 N_c^2 T^4\(1-\frac{15}{2}\g\)\(\chi(\hw,\hq)_{s,th} \frac{1+\frac{c_-}{c_+}\frac{Z''_{s,out}}{Z''_{s,in}}}{1+\frac{c_-}{c_+}\frac{Z_{s,out}}{Z_{s,in}}}\) \Bigg|_{u=0} \;,
\ea
where we have dropped the contact terms and $\chi_{s,th}$ is the thermal spectral density. All quantities have to be expanded consistently  to linear order in $\g$.
The relations between the retarded correlators of transverse stress, momentum density and energy density with the gauge invariant correlators are  given by \cite{Kovtun:2005ev}
\ba\label{Gretrel}
G_{xy,xy}&=&\frac{1}{2}G_1\;,\nn\\
G_{tx,tx}&=&\frac{1}{2}\frac{\hq^2}{\hw^2-\hq^2}G_2\;,\\
G_{tt,tt}&=&\frac{2}{3}\frac{\hq^4}{(\hw^2-\hq^2)^2}G_3\nn\;,
\ea
and the spectral density is defined as the  imaginary part of the retarded correlator  
\be
\chi_{\mu\nu,\rho\sigma}(\hw,\hq,u_s\g)=-2\, \textrm{Im} G_{\mu\nu,\rho\sigma}(\hw,\hq,u_s,\g).
\ee

To see how thermal equilibrium is approached it is instructive to look at  the relative deviation of the spectral density from its thermal  equilibrium value 
\be\label{R}
R_s(\hw,\hq,u_s,\g)=\frac{\chi_s(\hw,\hq,u_s,\g)-\chi_{s,th}(\hw,\hq,\g)}{\chi_{s,th}(\hw,\hq,\g)}.
\ee
This ratio is does not get altered by the relation for the symmetry channels with the retarded correlators (\ref{Gretrel}), therefore we also use the shorthand notation $\chi_s=-2 \,\textrm{Im}G_s$.

\section{Results}\label{results}
After describing the main parts of our computation we are now ready to investigate numerically the corresponding results.
We will start by analyzing  the quasinormal mode spectrum of the different channels   obtained from the thermal correlators and investigate the flow of the poles as a function of the coupling.
After that we will use the collapsing shell model to analyze the spectral densities and their approach to thermal equilibrium at finite coupling.

\subsection{Quasinormal mode spectrum}

Quasinormal modes characterize the response of the system to infinitesimal  external perturbations and are the strong coupling equivalent to quasiparticles  and branch cuts at weak coupling \cite{LeBellac, Hartnoll:2005ju}.
 They are solutions to linearized fluctuations of some bulk field obeying  incoming boundary conditions at the horizon and Dirichlet boundary conditions at the boundary. They appear as poles of the  corresponding retarded Green's function and have the generic form 
\be\label{frobenius}
\w_n(q)=M_n(q)-i\Gamma_n(q),
\ee 
 where $q$ is  the three momentum of the mode, $M_n$ and $\Gamma_n$ correspond to the mass and the decay rate of the excitation, respectively. 

\begin{figure*}[t]
\centering
\includegraphics[width=7.1cm]{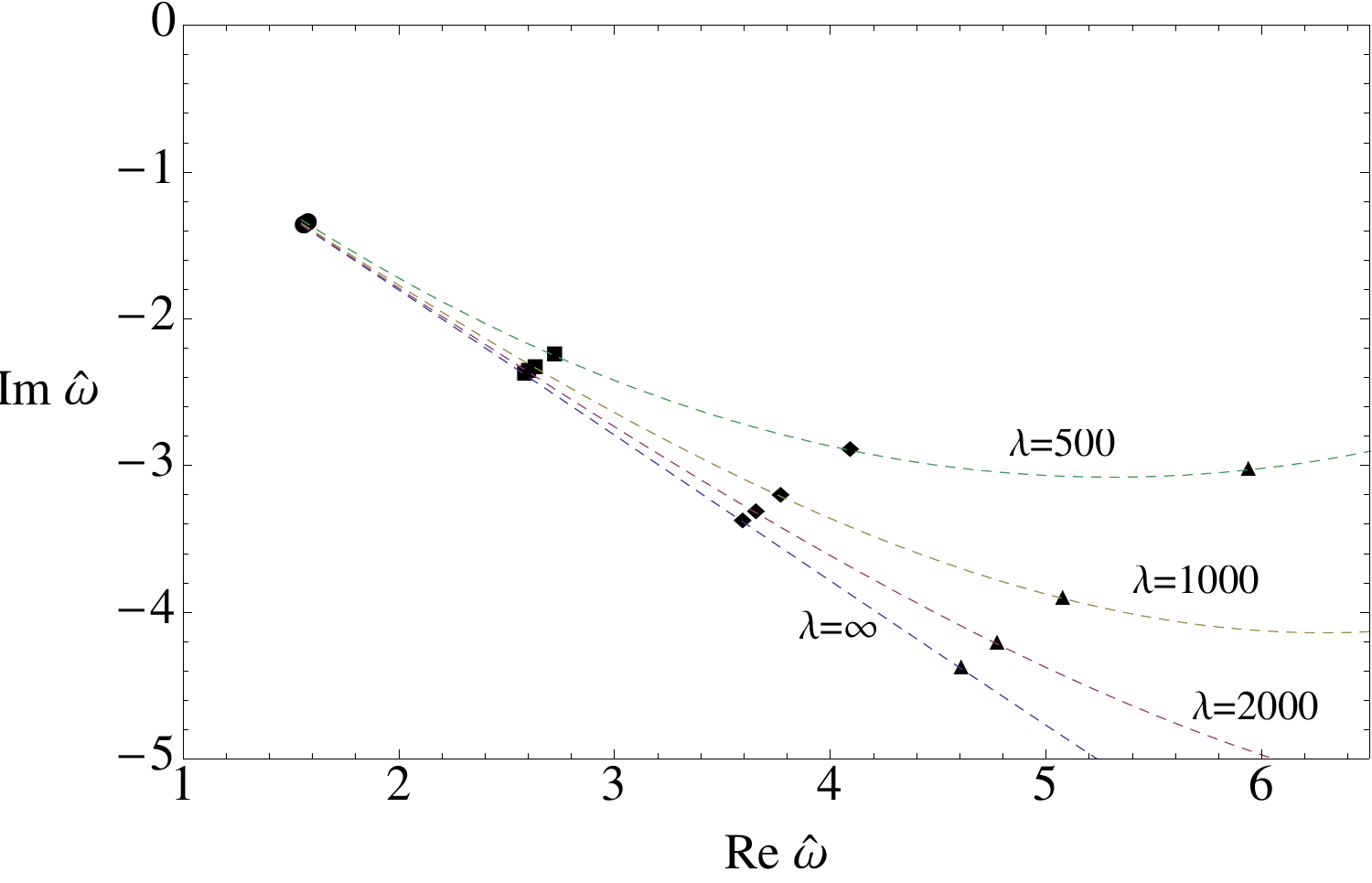}$\;\;\;\;\;\;\;\;$\includegraphics[width=7.1cm]{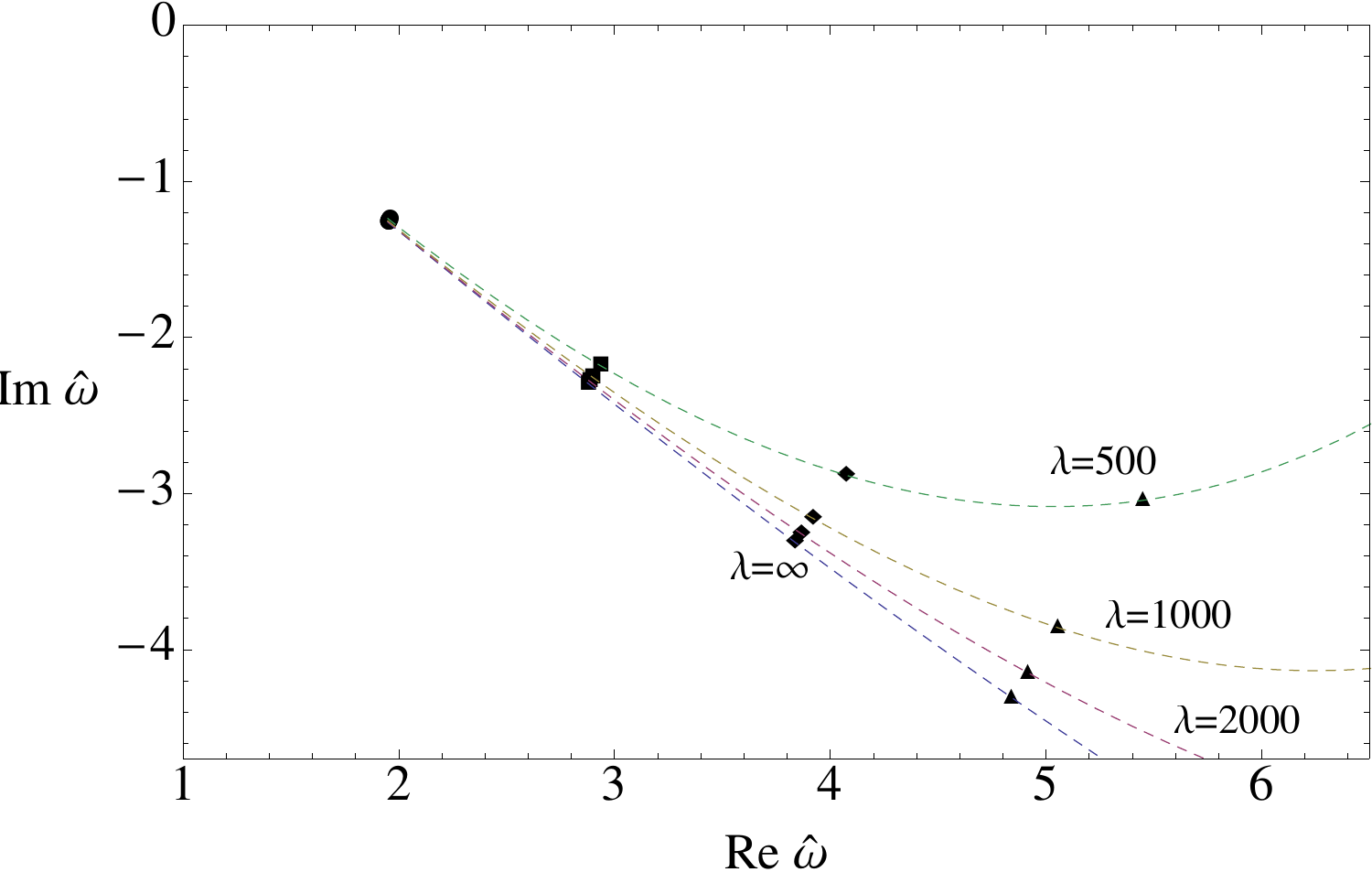}
\caption {The flow of the QNM in the scalar  channel for $q=0$ (left) and $q=2\pi T$ (right) as a function of $\lambda$.  The dashed lines are drawn to illustrate the bending of the tower of QNM towards the real axis as the coupling constant decreases. }
\label{QNMscalar}
\end{figure*}
\begin{figure*}[t]
\centering
\includegraphics[width=7.1cm]{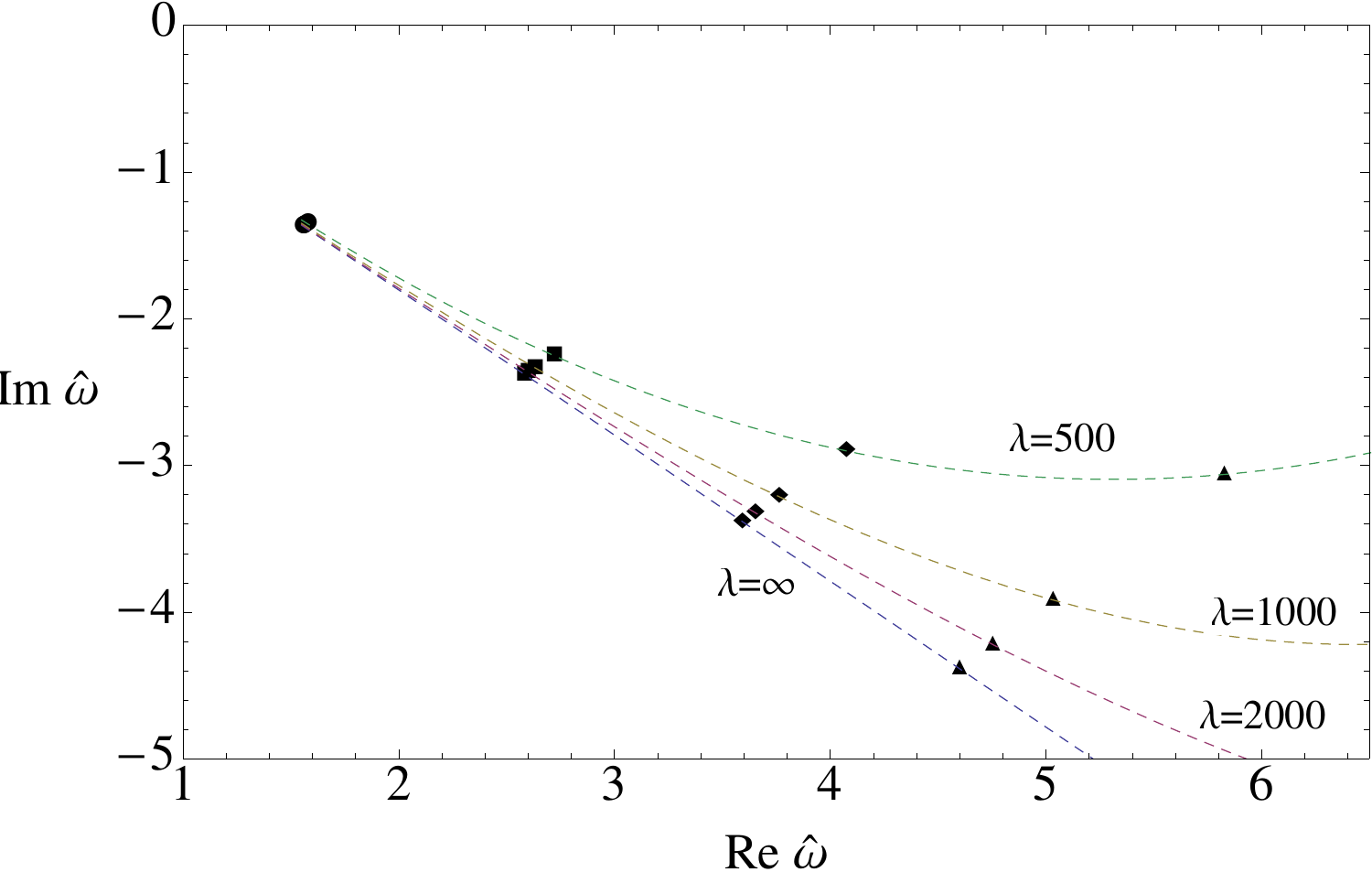}$\;\;\;\;\;\;\;\;$\includegraphics[width=7.1cm]{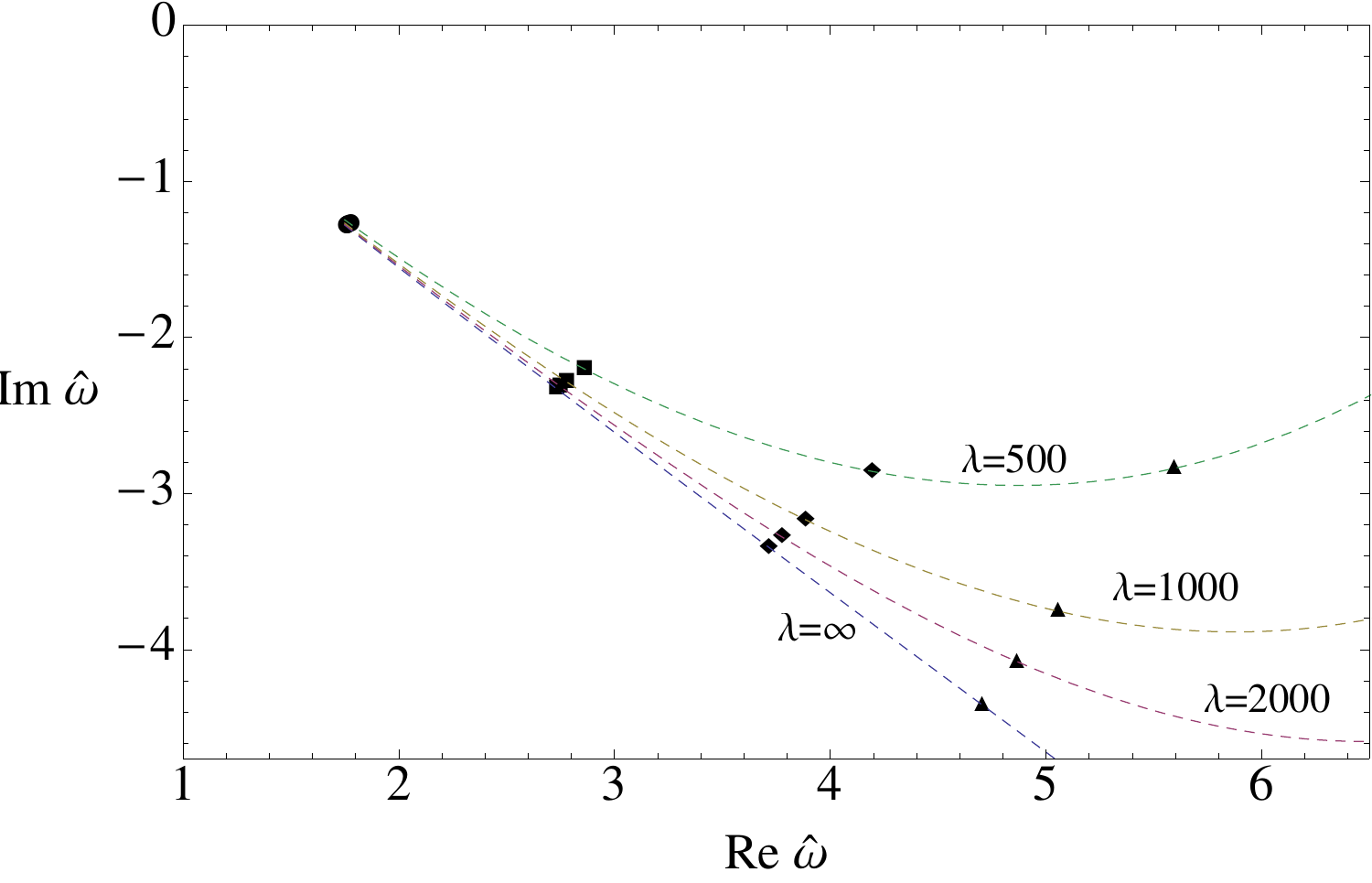}
\caption {The flow of the QNM in the shear channel for $q=0$ (left) and $q=2\pi T$ (right).}
\label{QNMshear}
\end{figure*}
\begin{figure*}[h]
\centering
\includegraphics[width=7.1cm]{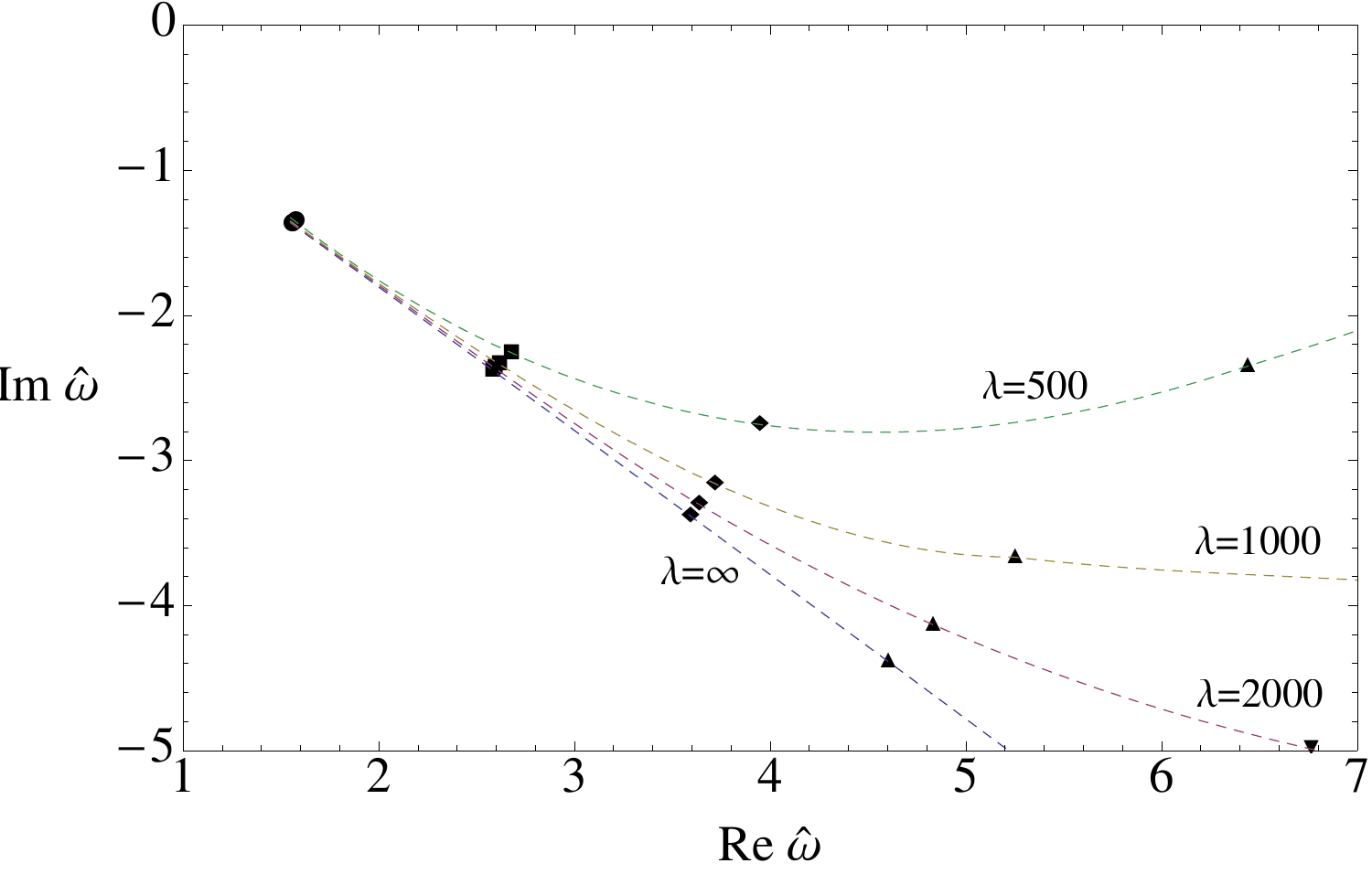}$\;\;\;\;\;\;\;\;$\includegraphics[width=7.1cm]{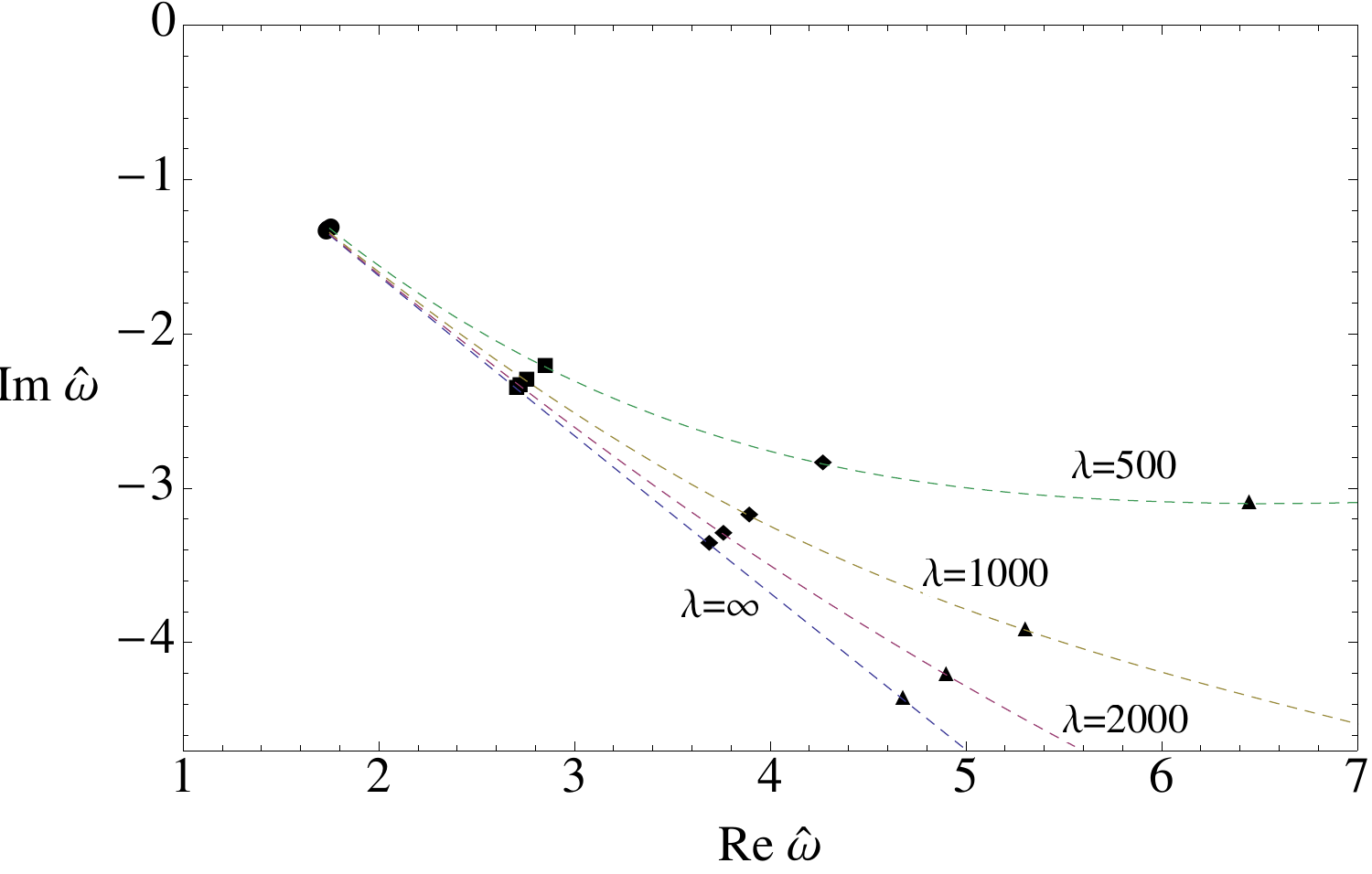}
\caption {The flow of the QNM in the sound channel for $q=0$ (left) and $q=2\pi T$ (right). }
\label{QNMsound}
\end{figure*}
For gravitational perturbations the QNM spectrum was first obtained in the infinite coupling limit in \cite{Kovtun:2005ev} and the diffusion poles in the hydrodynamic limit at finite coupling were worked out in
 \cite{Benincasa:2005qc, Buchel:2008bz}.
We are extending the existing analysis and study the flow of the tower of QNM  obtained in \cite{Kovtun:2005ev} as a function of the 't Hooft coupling $\lambda$.

In order to solve for the QNM spectrum we make a Frobenius ansatz for the ingoing modes
\be
Z_{s,in}^{(0)}(u,\w)=\sum_{n=0}^N a_{s,n}(\w)(1-u)^n\;, \qquad Z_{s,in}^{(1)}(u,\w)=\sum_{n=0}^N b_{s,n}(\w)(1-u)^n\;,
\ee
and solve recursively for the coefficients $a_{s,n}$ and $b_{s,n}$ by plugging the expansion into the equations of motion (\ref{eomscalar}), (\ref{eomsound}), (\ref{eomshear}), while the parameter $N$ is chosen large enough such that the behaviour of both functions is stable. 
We then make the ansatz for the frequencies, $\w=\w_0+\g \w_1$ and solve  numerically for the zeros of $Z_{s,in}(0,\w)=Z_{s,in}^{(0)}(0,\w)+ \g Z_{s,in}^{(1)}(0,\w)=0$.

The results for the flow of the QNMs is  displayed in fig.~\ref{QNMscalar} for the scalar, in  fig.~\ref{QNMshear} for the shear and in fig.~\ref{QNMsound} for the sound channel.
In all channels a clear trend is visible. As the coupling constant is lowered from the $\lambda=\infty$ limit
the imaginary part of $\w_n$ increases rapidly, lowering the decay rate of the excitation.
There is also a strong dependence on the index $n$: Higher energetic excitations show a stronger dependence on the coupling with a larger shift towards the real axis. These results point towards a weakening of the usual top-down thermalization pattern,  in accordance with the results found in \cite{Steineder:2013ana}, where an equivalent  calculation for the R-current correlator was performed. 
It should be noted, though,  that the strong coupling expansion can only be trusted fully if the relative deviation of the poles from the $\lambda=\infty$ limit is small, which clearly is not the case for all displayed poles.

\subsection{Thermalization of the spectral density}

Next the behaviour of the spectral density and its deviation from the thermal limit in the collapsing shell model is investigated.
To this end we parametrize the momentum of the plasma constituents by $q=c\,\w$. For $c=0$ the constituents of the plasma are at rest, while for $c=1$ they move with the speed of light.

\begin{figure*}[t]
\centering
\includegraphics[width=7.2cm]{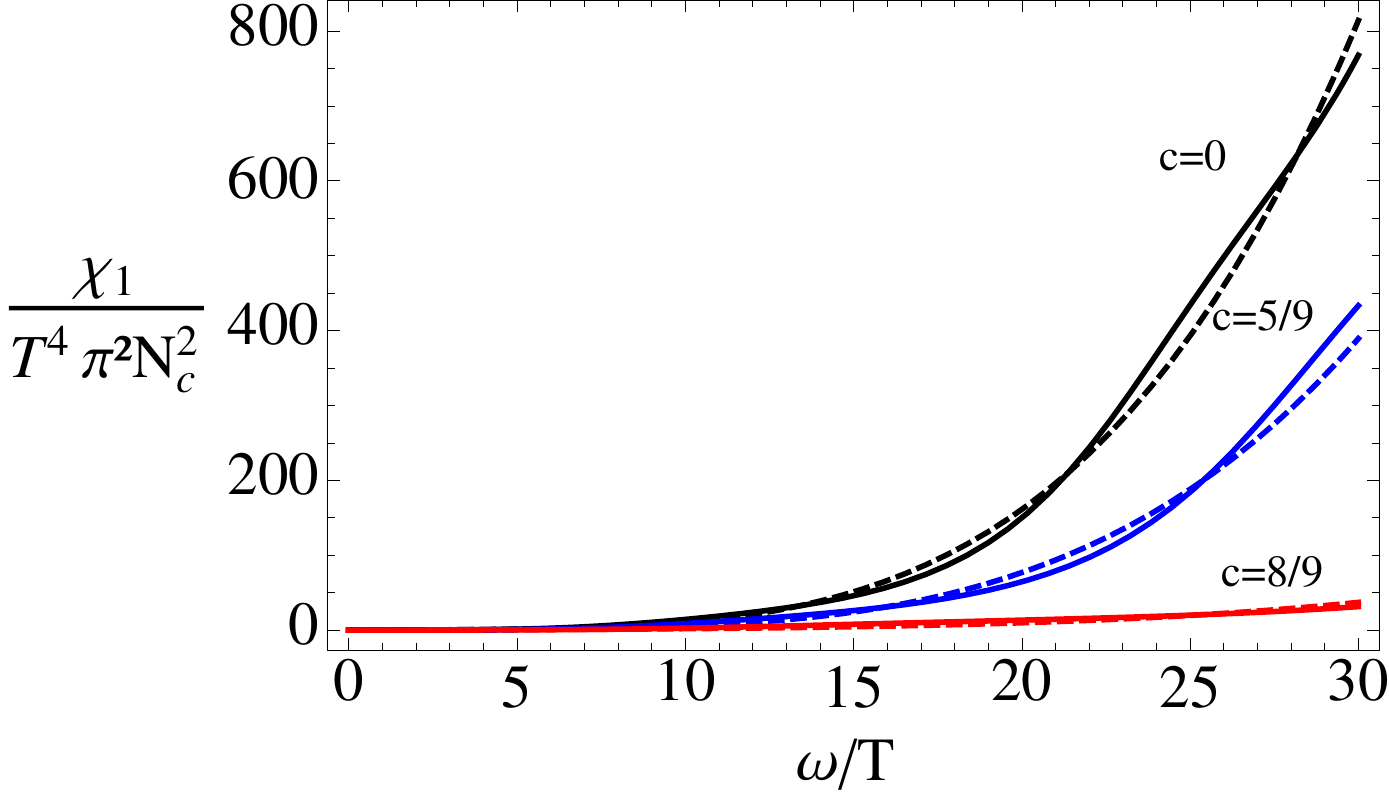}$\;\;\;\;\;\;\;\;$\includegraphics[width=7.2cm]{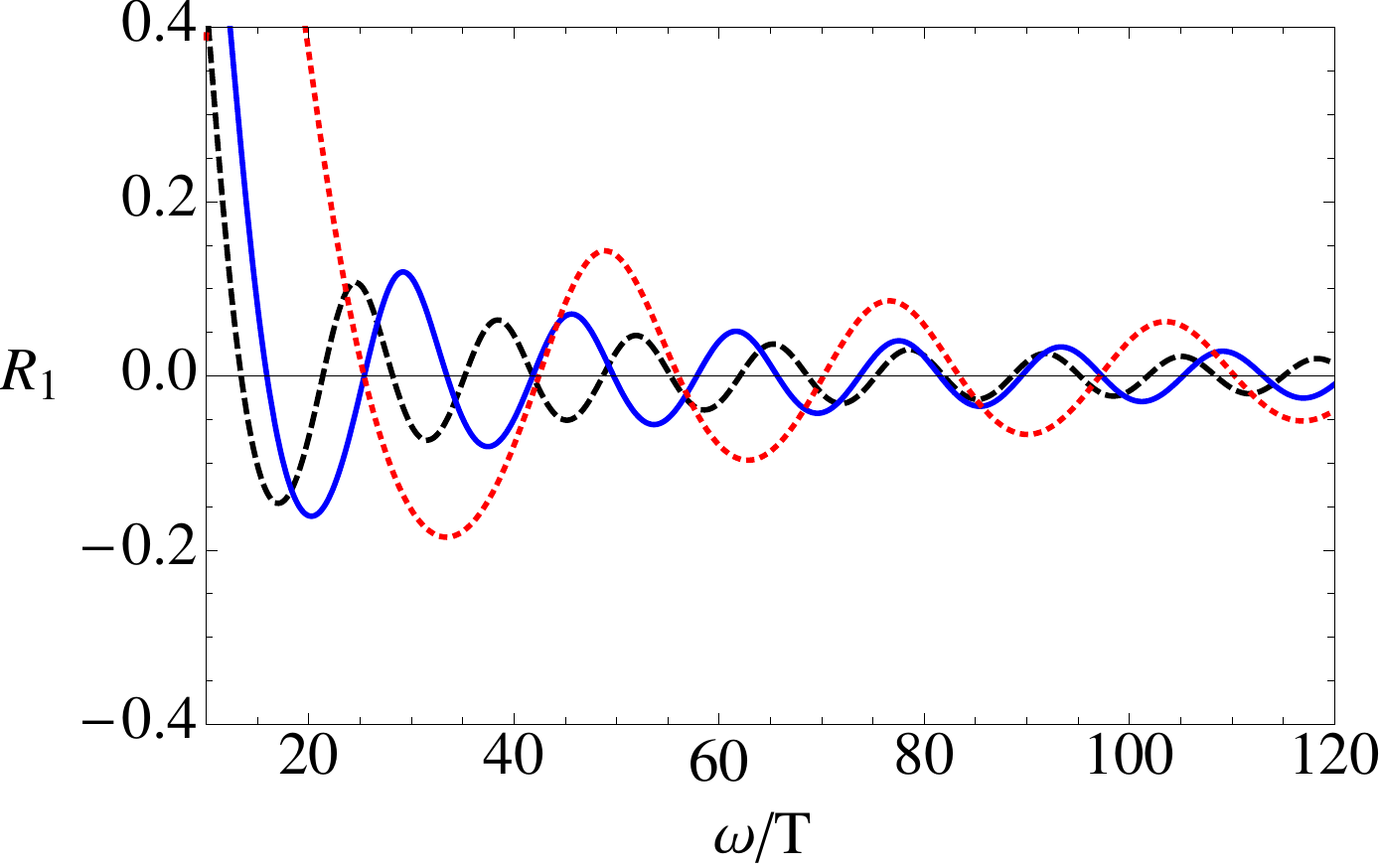}
\caption {Left: The spectral density for $\lambda=\infty$  in equilibrium (dotted lines) and out of equilibrium for $u_s=0.5$ .
Right: The relative deviation of the spectral density for $\lambda=\infty$, $c=8/9,\;5/9,\;0$ (from large to small amplitudes) and $u_s=0.5$ .
}
\label{specscalar}
\end{figure*}

In fig.~\ref{specscalar} we show the scalar  spectral density and its relative deviation  in and off-equilibrium in the infinite coupling limit for different values of c.
We witness oscillations of the off-equilibrium spectral densities around their equilibrium values and as the shell approaches the horizon the amplitude of the oscillations decreases\footnote{Since we are working in the quasi-static approximation (by  taking only  snapshots of the shell) and the effects of the shell location are minor, we set the shell location in all our plots to the rather arbitrary value $u_s=1/2$.}. 
From this figure one can also see that high energetic modes are closer to equilibrium than the low energetic ones, showing the usual top-down thermalization pattern with a  dependence on the parameter $c$. 
The smaller the value of $c$ the closer the quantity $R$ is to its equilibrium value.
\begin{figure}[t]
\centering
\includegraphics[width=7.2cm]{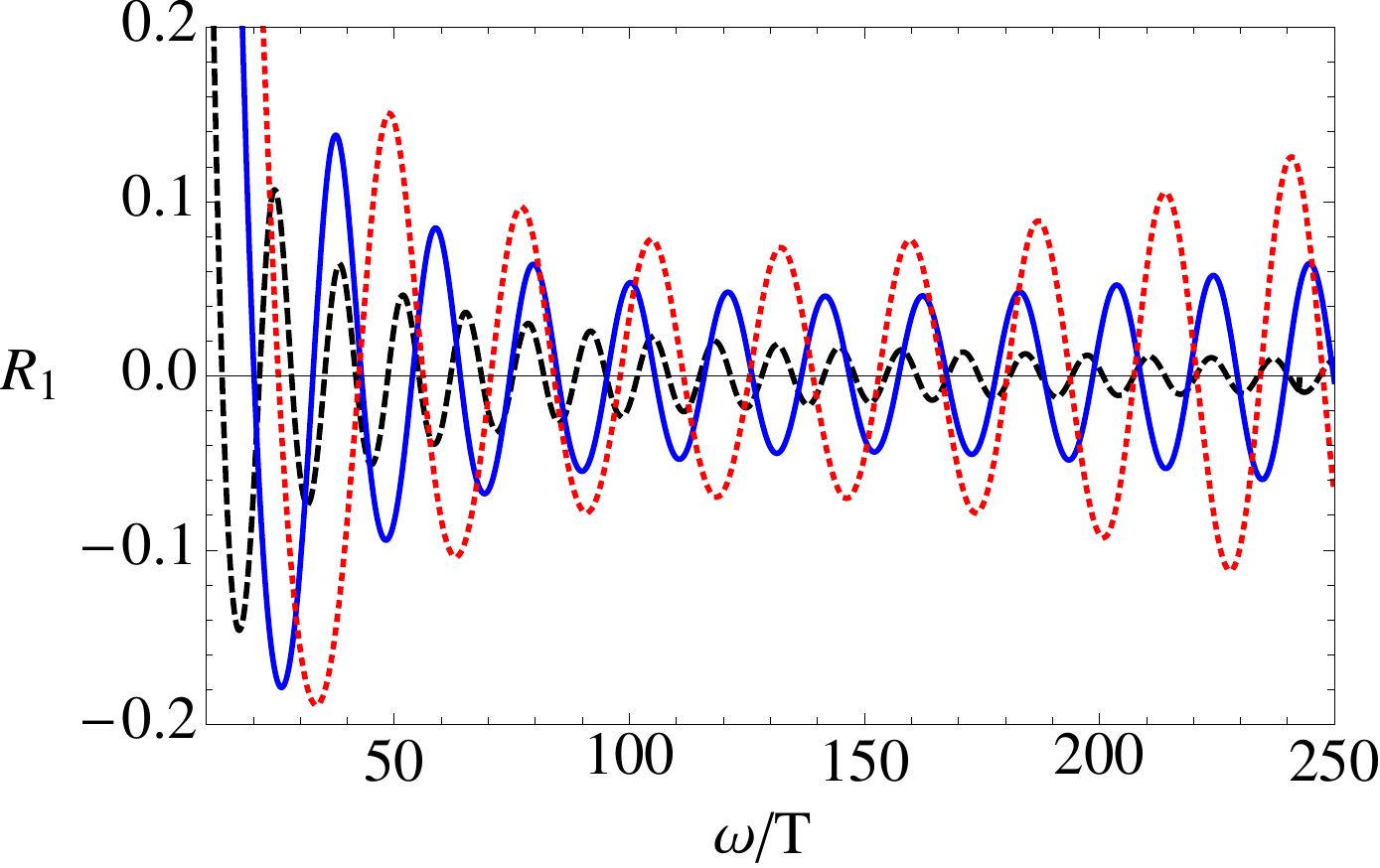}$\;\;\;\;\;\;\;\;$\includegraphics[width=7.2cm]{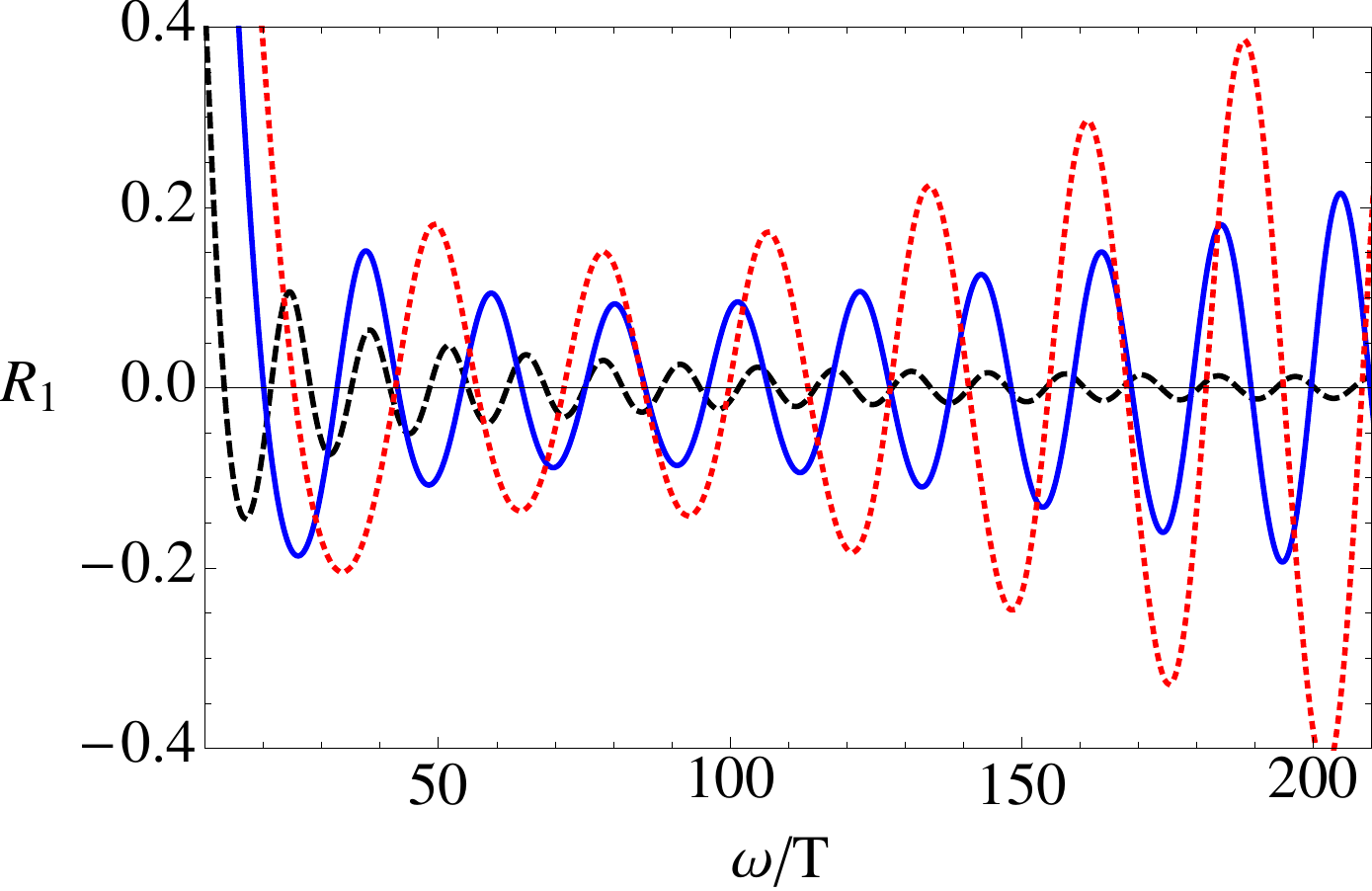}
\caption {The relative deviation of the spectral density, $R_1$, in the scalar  channel, for $c=0$ (dashed black), $c=7/9$ (solid blue), $c=8/9$ (dotted red), with the shell positioned at $u_s=0.5$ and $\lambda=300$ (left), $\lambda=100$ (right).}
\label{Rscalar}
\end{figure}

\begin{figure*}[t]
\centering
\includegraphics[width=7.2cm]{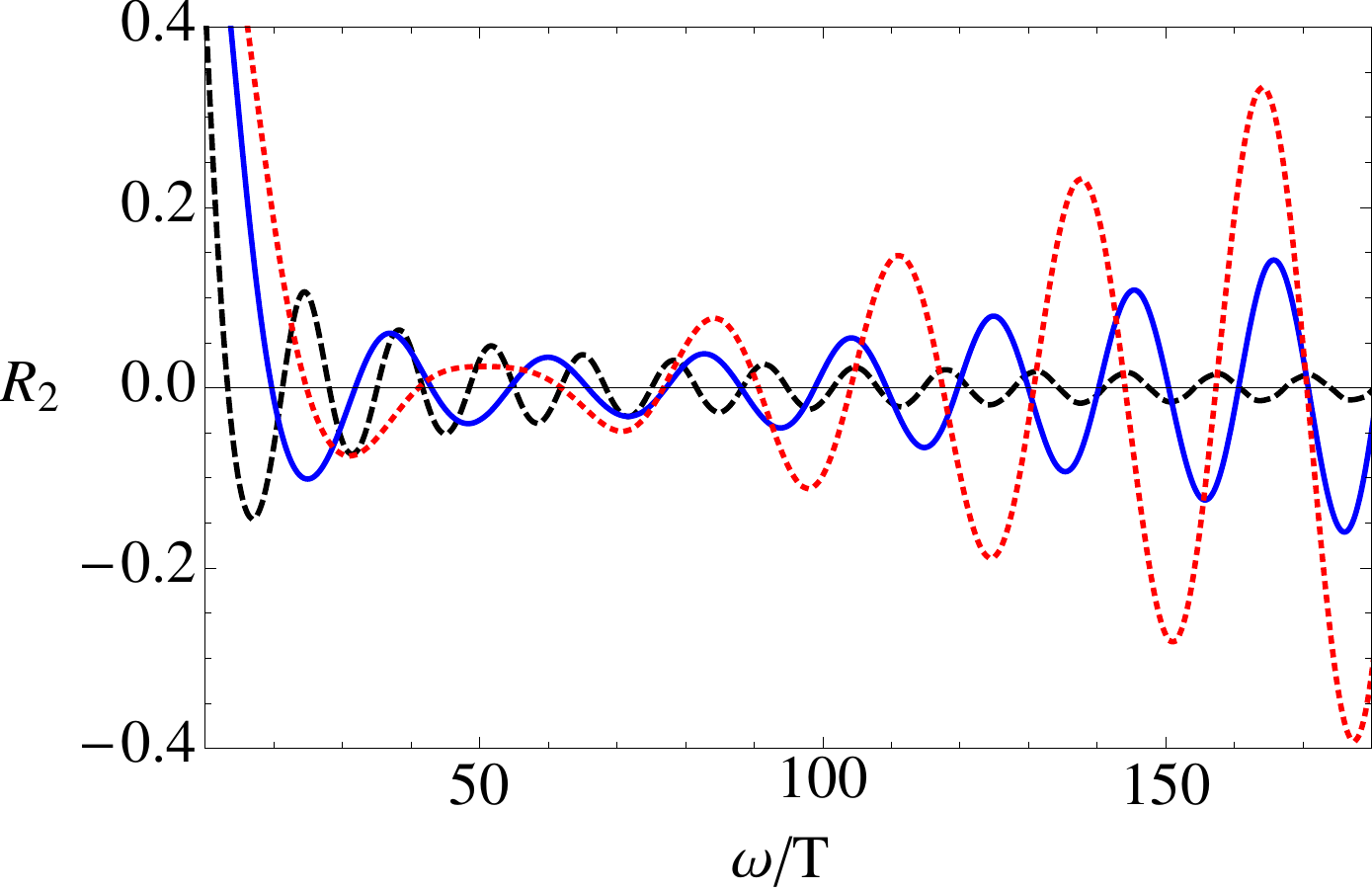}$\;\;\;\;\;\;\;\;$\includegraphics[width=7.2cm]{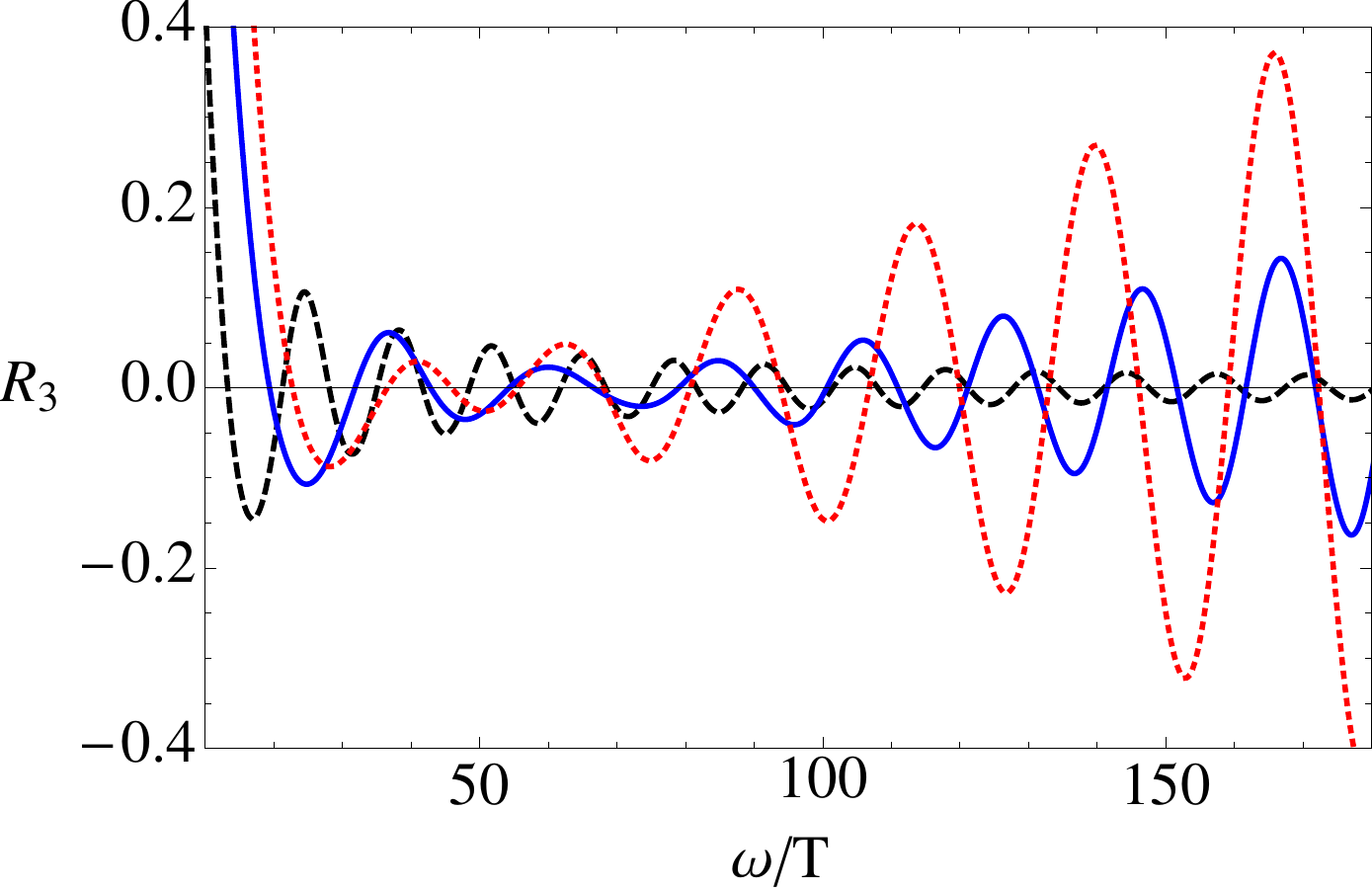}
\caption {The relative deviation of the spectral density  in the shear   channel (left) and sound channel (right) for  $\lambda=100$ and $u_s=0.5$. The colour coding is the same as in fig.~\ref{Rscalar}.}
\label{Rs}
\end{figure*}

Now we are ready to investigate the finite  coupling corrections to the relative deviation of the spectral density.
In fig.~\ref{Rscalar} the quantity $R$ is displayed for the scalar  channel for two different values of the coupling constant for a fixed position of the shell.
For plasma constituents  at rest, $c=0$, $R$ approaches a constant for large frequencies.
But as  $c$ is increased, the fluctuation amplitude starts to  grow at some  critical value of the frequency $\w_{crit}$, such that the higher energetic modes are further away from their equilibrium value than the low energetic ones,
 again indicating a  weakening of the usual top-down thermalization pattern. 
This fits nicely into the picture obtained for the QNM where also the higher energetic modes show a stronger  dependence on the finite coupling corrections.
As the coupling constant is decreased the change of the behaviour shifts to lower $\w_{crit}$. 
The results are again in accordance with the calculation for the spectral density of the R-current correlator \cite{Steineder:2013ana}.
Since the dependence of the coupling constant is qualitatively the same in the shear and sound channel, they are  only shown for one value of the coupling constant in  fig.~\ref{Rs}.

\begin{figure*}[h]
\centering
\includegraphics[width=7.8cm]{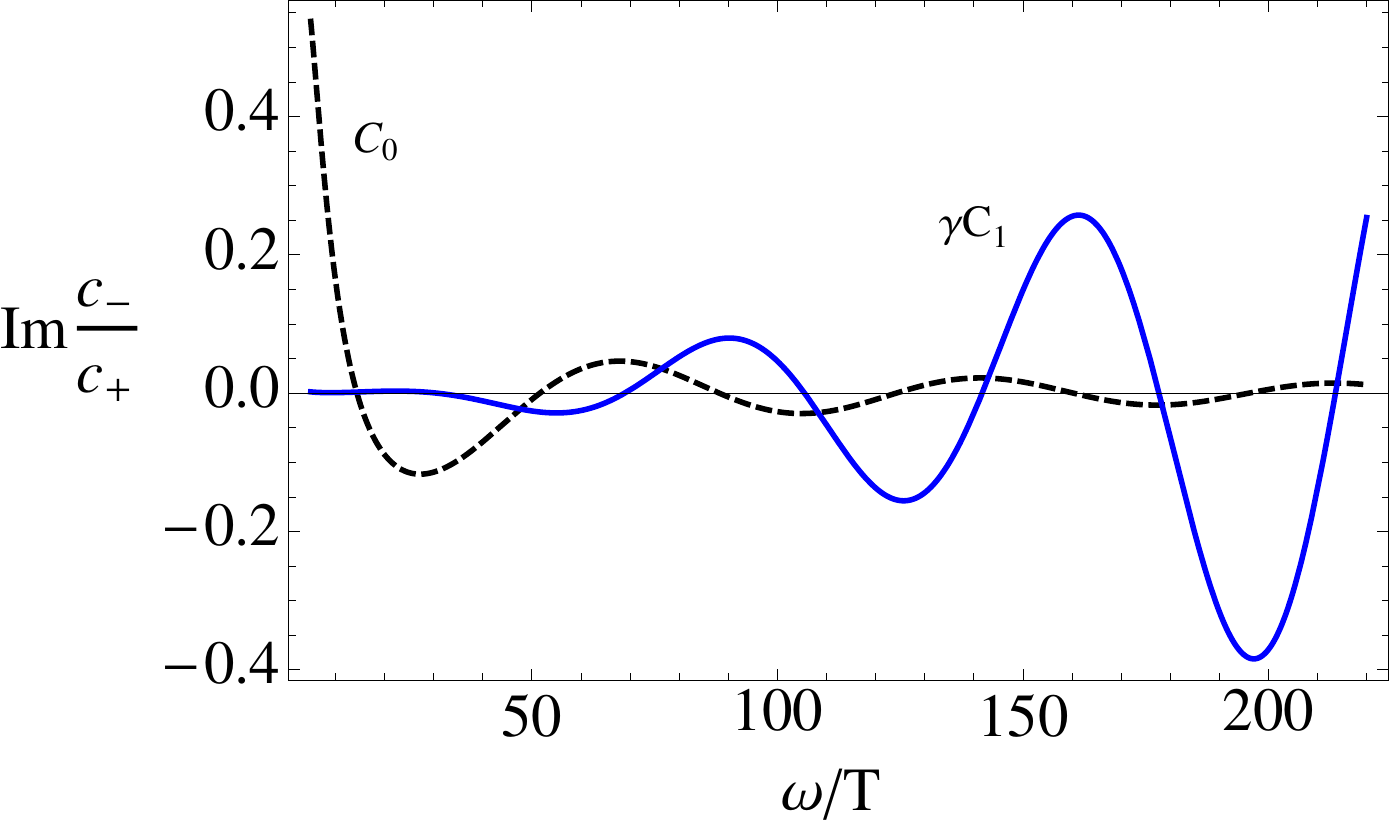}
\caption {Imaginary part of $C_0$ (dashed black) and $\gamma(\lambda=300)C_1$ (solid blue) as a function of frequency at $u_s=0.5$ and $c=8/9$. }
\label{cmp}
\end{figure*}

The parametric change of the relative deviation of the spectral density at finite coupling originates  from the behaviour of $C_0$ and $C_1$ as defined in  (\ref{Cmp}). As can be seen from fig.~\ref{cmp}, $C_0$ always approaches zero for large frequencies being responsible for the top-down thermalization pattern at infinite coupling.
On the other hand, the amplitude of $\gamma C_1$ is  constant for vanishing $c$ 
and starts growing as  $c$ is increased.
It is the interplay between $C_0$ and $C_1$ that is responsible for the observed pattern in $R$.

One might be worried about the above results since the quasi-static approximation and the finite coupling expansion employed have  finite regions of applicability. However, since we are only taking snapshots, and not treating the dynamical problem, one can few the system as being very close to its initial evolution where the motion of the shell is guaranteed to be slow. After all, we are allowed to inject energy at an arbitrary scale and set the initial velocity of the shell to zero.
In addition, from fig.~\ref{cmp} one might conclude that the strong coupling expansion breaks down once $\gamma C_1$ becomes larger than $C_0$. This however is not the case since the finite coupling corrections to the spectral density, which is the physical quantity,  are at most of the order of 10\%.

\section{Conclusions}\label{conclusion}

In the paper at hand we have studied the thermalization properties  of an $\mathcal{N}$=4 SYM plasma  at finite 't Hooft coupling.
First we analyzed how the plasma reacts to linearized perturbations as a function of the coupling constant through a QNM analysis.
Then we studied the approach of the plasma to thermal equilibrium  using the collapsing shell model of 
\cite{Danielsson:1999zt},  working in the quasi-static approximation.

The flow of the QNM is depicted in figs.~\ref{QNMscalar}, \ref{QNMshear}, \ref{QNMsound} and show a clear trend.
As the coupling constant is decreased from the $\lambda=\infty$ limit the QNM bend upwards in the complex frequency plane. The increase of the imaginary part shows, according to equ. (\ref{frobenius}),  that  finite coupling corrections increase the life time of the excitations. In addition,   higher energetic modes show a stronger dependence on the coupling corrections. We interpret this as a weakening of the usual top-down thermalization pattern.
 This is in  accordance with \cite{Steineder:2013ana} where a similar  analysis was performed for virtual photons. 
This analysis is particularly  useful because it is independent of the thermalization model being used and should therefore be of more general validity.

In the collapsing shell model, the deviation of the spectral density from its thermal limit was investigated. The results displayed in figs.~\ref{Rscalar} and \ref{Rs} 
show that outside the limit of infinite coupling, the UV modes are  further away from their thermal distribution than the IR modes, indicating a weakening of the top-down thermalization pattern.
Both the spectral density and the flow of the quasinormal modes show qualitatively the same pattern as  observed for real and virtual photons in \cite{Steineder:2013ana}, suggesting  that the change in the thermalization pattern is of more general validity.

The above observations seem, however, to be in direct contradiction with a recent study of thermalization at finite coupling using the Vaidya metric \cite{Baron:2013cya}, where  the UV modes were seen to thermalize even slightly faster than in the infinite coupling limit.
One possible explanation for  the discrepancy between their work and the calculation for photons presented in \cite{Steineder:2013ana} is an additional contribution of the Ramond-Ramond five form field strength that has to be added to the action  if photons are considered \cite{Paulos:2008tn}.
The analysis presented in this paper shows that the spectral density of the 
energy momentum tensor, to which this term does not contribute \footnote{In \cite{Myers:2008yi} it was shown that the Ramond-Ramond five form does not contribute to the equations of motion in the shear channel and therefore does not effect the spectral density.
It is expected that the same also holds for the other symmetry channels.}, exhibits the same behaviour as the spectral density for photons. Therefore the additional term for photons is not the source of the discrepancy.
Another important difference between the calculation of \cite{Baron:2013cya} 
and the  present one, as well as  the one of \cite{Steineder:2013ana}, is that the latter two use the
quasi-static approximation, i.e.~work in the limit of a slowly moving shell, while the first employs the opposite Vaidya limit. In addition, the correlation functions studied in \cite{Baron:2013cya} are all so-called geometric probes, meaning that they are only sensitive to the $\gamma$-corrected metric and
one need not consider fluctuation equations. Which of these differences
explain(s) the observed results remains, however, an open question, and a
very important topic for further investigation.

For other future directions, it will be important to go beyond the case studied here, 
 where we only take  snapshots of the system, and study the time evolution of the correlators within the quasi-static approximation along the lines of \cite{Lin:2013sga}. Of course, in the long run the goal is to consider finite coupling corrections to spectral densities in a thermalizing system without using the quasi-static approximation at all.

\section*{Acknowledgements}

The author would like to  thank R. Baier, V. Ker\"anen, H. Nishimura, A. Rebhan, D. Steineder and  A. Vuorinen for valuable discussion.
This work was supported by the START project Y435-N16 of the Austrian Science
Fund (FWF) and by the Sofja Kovalevskaja program of the Alexander von Humboldt foundation.
The author would also like to thank the Holograv network for financial support.

\begin{appendix}
\section{Quasi-static approximation}\label{A}
Here we are going to repeat the analysis of the applicability of the quasi-static approximation of \cite{Lin:2013sga} for the metric perturbation $h_{xy}$.
The induced metric on the shell can be put into the form
\be
ds_\Sigma^2=\frac{-d\tau^2+d\vec{x}^2}{(z(\tau))^2},
\ee
where $z(\tau)$ is the position of the shell at some world sheet time $\tau$.
Parametrizing $(t_\pm,z)=(t_\pm (\tau),z(\tau))$, where $u=z^2/z_h^2$ it follows from the Israel matching conditions that \cite{Lin:2013sga}
\be\label{t}
\dot{t}_-=\sqrt{1+\dot{z}^2}\,,\qquad \dot{t}_+=\frac{\sqrt{f+\dot{z}^2}}{f},
\ee 
where $\cdot=d/d\tau$.
The matching conditions for the metric perturbations $h_{xy}$ (see also \cite{Lin:2008rw}) are
\be
\frac{\dot{z}}{f}\partial_t h_{xy}^++f\dot{t}\partial_z h_{xy}^+=\dot{z}\partial_t h^-_{xy}+\dot{t} \partial_z h_{xy}^-\;,
\ee
where all quantities are evaluated at the shells position.
In order for the quasi-static approximation  to hold the inequalities must be satisfied
\ba
\frac{\dot{z}}{f}\omega h^+_{xy}(\w)&\ll& f\dot{t}\partial_z h_{xy}^+(\w)\\
\dot{z}\omega_- h^-_{xy}(\w_-)&\ll& \dot{t}\partial_z h_{xy}^-(w_-)-\frac{2\kappa_5^2p\, h_{xy}^-}{3z}.\label{2}
\ea
Using (\ref{mcxy}) and (\ref{t}) we obtain for the first equation
\be
\dot{z}\omega \ll \sqrt{f(f+\dot{z}^2)}\frac{\partial_z h_{xy}^-(\w/\sqrt{f_s})}{h_{xy}^-(\w/\sqrt{f_s})}.
\ee
Now making use of the analytic inside solution in terms of the Hankel function (\ref{hankel}) and $q=0$ the condition for the quasi-static approximation becomes
\be
\dot{z}\ll\(-\frac{2 H_0^{(2)}\(\frac{2 \w z}{\sqrt{f}}\)}{2 H_0^{(2)}\(\frac{2 \w z}{\sqrt{f}}\)}-\frac{1}{\w z}\)\sqrt{f(f+\dot{z}^2)}.
\ee
Similarly for (\ref{2}) we obtain
\be
\dot{z}\ll\(-\frac{2 H_0^{(2)}\(\frac{2 \w z}{\sqrt{f}}\)}{2 H_0^{(2)}\(\frac{2 \w z}{\sqrt{f}}\)}-\frac{1}{\w z}\)\sqrt{f(1+\dot{z}^2)}-\frac{2\kappa_5^2\,p}{3z^2 \w}f.
\ee

We will always place the shell at some initial position $z_i$, with vanishing initial velocity and then investigate the spectral density at positions of the shell close enough to its initial position and large enough frequencies  such that the quasi-static approximation holds. 
As discussed in  \cite{Lin:2013sga}, this type of an initial condition may actually not
be such a bad approximation to the initial conditions in a real life heavy
ion collision, as one can view $z_i$ to be in a rough correspondence with the
saturation scale $Q_s$.

\end{appendix}


\bibliographystyle{JHEP} 

\bibliography{refs.bib}

\end{document}